\documentstyle[12pt]{article}
\begin{document}
\begin{center}
{\Large\bf The Causal Phase in $QED_{3}$}
\vspace{1.0cm}
\\
J. L. Boldo\footnote{Supported by CAPES},
B. M. Pimentel\footnote{Partially supported by CNPq}
and J. L. Tomazelli\footnote{Supported by CAPES}
\vspace{0.2cm}

Instituto de F\'{\i}sica Te\'{o}rica \\
Universidade Estadual Paulista \\
Rua Pamplona, 145 \\
01405--900 - S\~{a}o Paulo, SP - Brazil

\vspace{1.0cm}

{\bf Abstract}
\end{center}
\vspace{0.3cm}

The operator ${\bf S}$ in Fock space which describes the scattering and
particle production processes in an external time-dependent electromagnetic
potential $A$ can be constructed from the one-particle S-matrix up to a
physical phase $\lambda [A]$. In this work we determine this phase for $QED$
in (2+1) dimensions, by means of causality, and show that no ultraviolet
divergences arise, in contrast to the usual formalism of $QED$.

\vspace{0.4cm}


\vspace{1.0cm}

{\large\bf 1.\,Introduction}
\vspace{1.0cm}

The efforts to test quantum electrodynamics in strong electromagnetic fields
in the late 70's brought into evidence the external field problem in the
context of the spontaneous dacay of the neutral to a charged vacuum through
pair creation, in heavy-ion collision experiments. Although the physics of
the quantized electron-positron field in interaction with a classical
electromagnetic field is well understood, some mathematical aspects of the
theory are rather involving, particularly the definition of the scattering
operator in Fock space for time-dependent external fields (for a review see
[1] and references therein).

In this paper we introduce the scattering operator ${\bf S}$ in Fock space
for quantum electrodynamics in (2+1) dimensional space-time, in an external
time-dependent electromagnetic field $A$ and show that it is unitary and
uniquely determined up to a phase. This phase is related to vacuum
fluctuations due to the presence of the external potential $A_{\mu}(x)$ and,
therefore, must depend on it. We then determine the phase $\lambda[A]$ in
lowest order of perturbation theory, by imposing Bogoliubov's local causality
condition on ${\bf S}$, and show that the vacuum-vacuum amplitude is
ultraviolet finite.

The construction of the S-matrix in Fock space is outlined in section 2.
In section 3 we present a brief digression on the global as well as the
differential causality conditions for the S-operator in Fock space. Section 4
is devoted to the derivation of the causal phase for $QED_3$ in lowest order
of perturbation theory, applying the concepts introduced in the preceeding
section, and exploiting the connnection with vacuum polarization. In section
5 we summarize our conclusions.

\vspace{1.0cm}

{\large\bf 2.\,The Scattering Operator in Fock Space}
\vspace{1.0cm}

We start from the one-particle Hamiltonian
\begin{equation}
H(t)=H_0+V(t) \,\,,
\end{equation}
where
\begin{equation}
V(t)=e(V(t,\vec{x})- \vec{\alpha}.\vec{A}(t,\vec{x})) \,\,.
\end{equation}
The potentials are assumed to vanish for $t \longrightarrow \pm \infty$ in
such a way that the wave operators
\begin{equation}
W_{\stackrel{in}{out}}=s-\lim_{t \rightarrow \pm
\infty}U(t,0)^{\dagger}e^{-iH_0t}
\end{equation}
exist, together with a unitary S-matrix
\begin{equation}
S=W_{out}^{\dagger}W_{in} \,\,.
\end{equation}

Since by assumption we have the free dynamics for $t \longrightarrow \pm
\infty$, we settle second quantization on the Fock representation of the free
Dirac field
\begin{equation}
\psi (f)=b(P^0_{+}f)+d(P^0_{-}f)^{\dagger} \,\,.
\end{equation}
Here $P^0_{\pm}$ are the projection operators on the positive and negative
spectral subspaces of the one-particle free Dirac Hamiltonian $H_0$,
respectively.

The second quantized S-matrix in Fock space is now defined by
\begin{equation}
\psi(S^{\dagger}f)={\bf S}^{-1}\psi(f){\bf S} \,\,,
\end{equation}
\begin{equation}
\psi(S^{\dagger}f)^{\dagger}={\bf S}^{-1}\psi(f)^{\dagger}{\bf S} \,\,,\,\,
\forall f \in {\cal H}_1 \,\,,
\end{equation}
if it exists. We have taken the adjoint $S^{\dagger}$ in the test functions
since $\psi (f)$ is antilinear in $f$. It follows from the above definitions
that ${\bf S}$ is unitary and uniquely determined up to a phase. In order to
prove this assertion we proceed as in reference $[2]$.

\vspace{0.5cm}

{\bf Proposition:} {\bf S} is uniquely determined by (6) and (7) up to a
factor.
\vspace{0.5cm}

{\em Proof.} If $\tilde{\bf S}$ is another operator in ${\cal F}$, satisfying
(6), then
 \[{\bf S}^{-1}\psi(f){\bf S}=\tilde{\bf S}^{-1}\psi(f)\tilde{\bf S} \,\,,\]
\[\tilde{\bf S}{\bf S}^{-1}\psi(f)=\psi(f)\tilde{\bf S}{\bf S}^{-1} \,\,,\,\,
\forall f \in {\cal H}_1 \,\,,\]
and the same is true for all $\psi^{\dagger}(f)$. From the irreducibility of
the Fock representation, we have
\begin{equation}
\tilde{\bf S}{\bf S}^{-1}=\alpha {\bf 1}\,\,\,\,{\rm i.e.}\,\,\,\,
\tilde{\bf S}=\alpha{\bf S} \,\,.
\end{equation}
\begin{flushright}
$\Box$
\end{flushright}

Now, taking the adjoint of (6)
\[\psi(S^{\dagger}f)^{\dagger}={\bf S}^{\dagger}\psi(f)^{\dagger}
{{\bf S}^{-1}}^{\dagger}  \,\,,\]
and comparing with (7), it follows again from the irreducibility of the Fock
representation that
\begin{equation}
{\bf S}^{\dagger}=\rho{\bf S}^{-1} \,\,.
\end{equation}

If we take the adjoint and the inverse of this equation, namely
\[{\bf S}=\rho^{*}{{\bf S}^{-1}}^{\dagger} \,\,,\]
\[{{\bf S}^{\dagger}}^{-1}=\rho^{-1}{\bf S} \,\,,\]
we find that
\[\rho^{*}=\rho \,\,.\]
{}From (8) and (9) we obtain
\begin{equation}
\tilde{\bf S}^{\dagger}=|\alpha|^2\rho \tilde{\bf S}^{-1} \,\,.
\end{equation}
Therefore, we may choose
\begin{equation}
|\alpha|^2=\rho^{-1}
\end{equation}
such that the operator $\tilde{\bf S}$ becomes unitary. Since the absolute
value of $\alpha$ in (8) is fixed by (11), $\tilde{\bf S}$ is uniquely
determined up to a phase $e^{i\lambda}$. However, this phase $\lambda[A]$ is
physical because it depends on the external potential $A_{\mu}(x)$. As we
shall see this phase will be fixed by the requirement of causality of
${\bf S}$.

The S-matrix ${\bf S}$ in Fock space exists, if and only if $P_{+}SP_{-}$ is
a Hilbert-Schmidt operator. In this case it is given by
\begin{equation}
{\bf S}=C\,e^{S_{+-}S_{--}^{-1}b^{\dagger}d^{\dagger}}
          :e^{(S_{++}^{\dagger -1}-1)b^{\dagger}b}:
	  :e^{(1-S_{--}^{-1})dd^{\dagger}}:
	  e^{S_{--}^{-1}S_{-+}db}\,\,,
\end{equation}
where
\begin{equation}
S_{ij}=P_iSP_j \,,\,\,\,i,j=+,-
\end{equation}
and
\begin{equation}
|C|^2={\rm det}(1-S_{+-}S_{-+}^{\dagger})\,\,.
\end{equation}

The first factor in (12) describes electron-positron pair creation, the
second one electron scattering, the third one positron scattering and the
last one pair annihilation.

\vspace{1.0cm}

{\large\bf 3.\,The Condition of Causality}
\vspace{1.0cm}

In the one-particle theory the condition that a change in the interaction law
in any space-time region can influence the evolution of the system only at
subsequent times can be translated into the factorization of the S-matrix
\begin{equation}
S[A]=S_2S_1\,\,,\,\,\,S_j \stackrel{def}{=}S[A_j]\,\,,
\end{equation}
where we have written the electromagnetic potential as
\begin{equation}
A^{\mu}(x)=A^{\mu}_1(x)+A^{\mu}_2(x)\,\,,
\end{equation}
which is the sum of two parts with disjoint supports in time
\begin{equation}
{\rm supp}\,A_1 \subset (-\infty,r]\,\,\,,\,\,\,{\rm supp}\,A_2 \subset
[r,+\infty)\,\,.
\end{equation}

A similar factorization should hold from Eq.(6) for the S-operator ${\bf S}$
in Fock space,
\begin{equation}
(\Omega,{\bf S}\Omega)=(\Omega,{\bf S_2S_1}\Omega)\,\,.
\end{equation}
We call (18) global causality condition for the Fock space S-operator in
contrast to the differential condition${}^{[4]}$
\begin{equation}
\frac{\delta}{\delta A_{\mu}(y)}\left(\Omega,{\bf S}^{\dagger}\frac{\delta {\bf
 S}}{\delta A_{\nu}(x)}\Omega\right)=0,\,\,\,{\rm for}\,\,\,
 x^0<y^0\,\,.
\end{equation}

We have seen in the last section that the S-matrix in Fock space can be
uniquely determined up to a phase,
\begin{equation}
{\bf S}=e^{i\varphi}\tilde{\bf S}\,\,,
\end{equation}
where $\tilde{\bf S}$ is unitary, and given by expression (12).
Inserting (20) into (19) we obtain
\begin{flushleft}
$\displaystyle{\frac{\delta}{\delta A_{\mu}(y)}\left({\bf S}\Omega,
\frac{\delta{\bf S}}{\delta A_{\nu}(x)}\Omega\right)}=$
\end{flushleft}
\begin{equation}
i\frac{\delta^2\varphi}{\delta A_{\mu}(y)\delta A_{\nu}(x)}+
\frac{\delta}{\delta A_{\mu}(y)}\left(\tilde{\bf S}\Omega,\frac
{\delta\tilde{\bf S}}{\delta A_{\nu}(x)}\Omega\right)\,\,.
\end{equation}

It can be shown from the unitarity of $\tilde{\bf S}$ that the last term in
(21) is purely imaginary. Consequently, the real part of the causality
condition (19) is automatically satisfied while for the imaginary part we may
choose $\varphi$ conveniently such that (19) holds.

\vspace{1.0cm}

{\bf 4.\,The Causal Phase}
\vspace{1.0cm}

We now turn to the determination of the causal phase in lowest order of
perturbation theory. From (6) we have
\begin{equation}
\tilde{\bf S}\Omega=C(\Omega+\sum_{mn}(S_{+-})_{mn}
b_{m}^{\dagger}d_{n}^{\dagger}\Omega+ \dots)\,\,,
\end{equation}
where we have put $S_{--}^{-1}$ equal to the unity in lowest order.
Taking the functional derivative of (22) with respect to $A_{\nu}(x)$ and
keeping only terms of order $O(A)$ in the resulting expression, we arrive at
\begin{equation}
\left(\tilde{\bf S}\Omega,\frac{\delta\tilde{\bf S}}{\delta A_{\nu}(x)}
\Omega\right)=iC^2\,\Im m {\rm Tr}\left(S_{-+}^{\dagger}\frac{\delta S_{+-}}
{\delta A_{\nu}(x)}\right)\,\,.
\end{equation}
In lowest order we may set $C^2=1$.

The local causality condition (19) together with expressions (21) and
(23) yield
\begin{equation}
F(x,y) \stackrel{def}{=} \frac{\delta^2\varphi}{\delta A_{\mu}(y)\delta
A_{\nu}(x)} +\Im m\frac{\delta}{\delta A_{\mu}(y)}{\rm Tr}\left({(S_{+-})}
^{\dagger}\frac{\delta S_{+-}}{\delta A_{\nu}(x)}\right)=0
\end{equation}
for $x^0<y^0$.

Next we calculate the second term in (24). In lowest order of perturbation
theory, we have
\begin{equation}
S_{+-}^{(1)}=-iP_{+}({\bf p})\gamma^0e\,\slash\!\!\!\!A(p+q)P_{-}(-{\bf q})
\,\,.
\end{equation}
As in reference $[3]$ we use the following representation for the Dirac
matrices in (2+1) dimensions:
\begin{equation}
\gamma^0=\sigma_3\,,\,\,\,\gamma^1=i\sigma_1\,,\,\,\,\gamma^2=i\sigma_2\,\,
\end{equation}
where $\sigma_j$ are the Pauli matrices.

{}From (25) we obtain
\begin{flushleft}
${\rm Tr}\displaystyle{\frac{\delta}{\delta A_{\mu}(y)}{(S_{+-})}^{\dagger}}
\displaystyle{\frac{\delta S_{+-}}{\delta A_{\nu}(x)}} =$

\vspace{0.3cm}

$\displaystyle{e^2{(2\pi)}^{-3}\int\,d^2p \int\,d^2q\,e^{i(p+q)(x-y)}
{\rm tr}[P_{-}(-{\bf q})\gamma^0\gamma^{\mu}P_{+}({\bf p})\gamma^0\gamma^
{\nu}P_{-}(-{\bf q})]}$
\end{flushleft}
\begin{equation}
=-\int\,d^3k\,e^{ik(x-y)}\hat{P}^{\mu \nu}(k)\,\,,
\end{equation}
$\hat{P}^{\mu \nu}(k)$ is not but the vacuum polarization tensor in
(2+1) dimensions, which is given by${}^{[3]}$
\begin{equation}
\hat{P}^{\mu \nu}(k)=-e^2{(2\pi)}^{-3}T^{\nu \mu}(k)\,\,,
\end{equation}
where
\begin{eqnarray}
T^{\nu \mu}(k) &   =  & \int\,d^3p\,\,\delta(p^2-m^2)\Theta(p^0)
\delta[(k-p)^2-m^2]\nonumber \\
               &\times& \Theta(k^0-p^0)\,t^{\nu \mu}(k,p)
\end{eqnarray}
with
\begin{equation}
t^{\nu \mu}(k,p)={\rm tr}[\gamma^{\mu}(\slash\!\!\!p+m)\gamma^{\nu}
(k\!\!\!\slash-\slash\!\!\!p-m)]\,\,.
\end{equation}

It follows from the gauge invariance of (28) that
\begin{equation}
\hat{P}^{\mu \nu}(k)=\hat{P}_S^{\mu \nu}(k)+\hat{P}_A^{\mu \nu}(k)
\end{equation}
with
\begin{eqnarray}
\hat{P}_S^{\mu \nu}(k)=(k^{\mu}k^{\nu}-k^2g^{\mu \nu})\tilde{B}(k^2)\,\,,\\
\nonumber \\
\hat{P}_A^{\mu \nu}(k)=im\varepsilon^{\mu \nu \alpha}k_{\alpha}\tilde{\Pi}
^{(2)}(k^2)\,\,.
\end{eqnarray}
Performing the trace in (30) and the resulting momentum integral in (29) we
find that${}^{[3]}$
\begin{eqnarray}
\tilde{B}(k^2)=\frac{-e^2}{2{(4\pi)}^2}\frac{k^2+4m^2}{k^2}\Theta(k^2-4m^2)
\frac{\Theta(k_0)}{\sqrt{k^2}}\,\,,\\
\nonumber \\
\tilde{\Pi}^{(2)}(k^2)=\frac{-e^2}{2{(2\pi)}^2}\Theta(k^2-4m^2)
\frac{\Theta(k_0)}{\sqrt{k^2}}\,\,.
\end{eqnarray}

Substituting (27) and (28) in (24), we rewrite the causal function $F(x,y)$
as
\begin{equation}
F(x,y) = \frac{\delta^2\varphi}{\delta A_{\mu}(y)\delta
A_{\nu}(x)} +\frac{e^2}{{(2\pi)}^3}\Im m\,\int\,d^3k\,e^{ik(x-y)}
T^{\nu \mu}(k)\,\,.
\end{equation}

We can evaluate the imaginary part of the last term in the above equation
taking into account (28) and (31)-(35). Thus, we have
\begin{equation}
T^{\nu \mu}(k)=T^{\nu \mu}_S(k)+T^{\nu \mu}_A(k)\,\,,
\end{equation}
where $T^{\nu \mu}_S(k)$ is real and even in k while $T^{\nu \mu}_A(k)$ is
imaginary and odd in k. Hence,
\begin{flushleft}
$F(x,y) = \displaystyle{\frac{\delta^2\varphi}{\delta A_{\mu}(y)\delta
A_{\nu}(x)}}+$
\end{flushleft}
\begin{equation}
\frac{e^2}{{(2\pi)}^3}\left[\int_{k_0>0}\,d^3k\,\sin
k(x-y)T^{\nu \mu}_S(k)-i\int_{k_0>0}\,d^3k\,\cos k(x-y)T^{\nu \mu}_A(k)
\right]\,\,.
\end{equation}
In order to write the last term in (38) as a complex Fourier transform we
must continue $T^{\nu \mu}(k)$ antisymmetrically to $k_0<0$
\begin{equation}
F(x,y) = \frac{\delta^2\varphi}{\delta A_{\mu}(y)\delta
A_{\nu}(x)} +\frac{i}{2}\int\,d^3k\,e^{-ik(x-y)}[d_S^{\mu \nu}(k)-
d_A^{\mu \nu}(k)]\,\,,
\end{equation}
where
\begin{eqnarray}
d_S^{\mu \nu}(k)=(k^{\mu}k^{\nu}-k^2g^{\mu \nu})B(k^2)\,\,,\\
\nonumber \\
d_A^{\mu \nu}(k)=im\varepsilon^{\mu \nu \alpha}k_{\alpha}\Pi
^{(2)}(k^2)\,\,.
\end{eqnarray}
and
\begin{eqnarray}
B(k^2)=\frac{-e^2}{2{(4\pi)}^2}\frac{k^2+4m^2}{k^2}\Theta(k^2-4m^2)
\frac{{\rm sgn}(k_0)}{\sqrt{k^2}}\,\,,\\
\nonumber \\
\Pi^{(2)}(k^2)=\frac{-e^2}{2{(2\pi)}^2}\Theta(k^2-4m^2)
\frac{{\rm sgn}(k_0)}{\sqrt{k^2}}\,\,.
\end{eqnarray}

The Fourier transform of a causal function vanishing for $x^0-y^0=t<0$
satisfies a dispersion relation. Since $d_S^{\mu \nu}(k)$ and
$d_A^{\mu \nu}(k)$ are real and purely imaginary, respectively, they cannot
be the Fourier transform of a causal function. The lacking imaginary part of
$d_S^{\mu \nu}(k)$ and the lacking real part of $d_A^{\mu \nu}(k)$ must de
supplied by the first term containing the phase $\varphi[A]$,
\begin{equation}
\frac{\delta^2\varphi}{\delta A_{\mu}(y)\delta
A_{\nu}(x)}=\frac{i}{2}\int\,d^3k\,e^{-ik(x-y)}[ir_S^{\mu \nu}(k)-
r_A^{\mu \nu}(k)]\,\,,
\end{equation}
where
\begin{flushleft}
$r_S^{\mu \nu}(k)=\displaystyle{\frac{1}{2\pi}\int_{-\infty}^{+\infty}\,dt
\frac{d_S^{\mu \nu}(kt)}{{(t-i0)}^2(1-t+i0)}}$
\end{flushleft}
\begin{equation}
=\frac{\beta}{2\pi}(k^{\mu}k^{\nu}-k^2g^{\mu \nu})\left[\frac
{1}{\sqrt{k^2}}\left(1+\frac{4m^2}{k^2}\right)\log \left(\frac{1-
\sqrt{\frac{4m^2}{k^2}}}{1+\sqrt{\frac{4m^2}{k^2}}}\right)+\frac{4m}{k^2}
\right]\,\,,
\end{equation}
and
\begin{flushleft}
$r_A^{\mu \nu}(k)=-\displaystyle{\frac{i}{2\pi}\int_{-\infty}^{+\infty}\,dt
\frac{d_A^{\mu \nu}(kt)}{(t-i0)(1-t+i0)}}$
\end{flushleft}
\begin{equation}
=\frac{i\beta}{2\pi}4im\varepsilon^{\mu \nu \alpha}\frac{k_\alpha}
{\sqrt{k^2}}\log \left(\frac{1-\sqrt{\frac{4m^2}{k^2}}}{1+\sqrt{\frac{4m^2}
{k^2}}}\right)\,\,,
\end{equation}
with $\beta \equiv -e^2/[2{(4\pi)}^2]$.

The causal phase is obtained by two integrations
\begin{flushleft}
$\varphi[A]=\displaystyle{\frac{1}{2}\int\,d^3x\,\int\,d^3y\,
\frac{\delta^2\varphi}{\delta A_{\mu}(y)\delta A_{\nu}(x)}A_{\mu}(y)A_{\nu}(x)
\,\,+O(A^4)}$
\end{flushleft}
 \begin{equation}
=-\pi^2\beta\int\,d^3k\left[\left(\frac{k^{\mu}k^{\nu}}{k^2}-g^{\mu
\nu}\right)\Pi_1^{(1)}(k)+im\varepsilon^{\mu \nu \alpha}k_{\alpha}
\Pi_1^{(2)}(k)\right]A_{\mu}(k)A_{\nu}^{*}(k)\,\,,
\end{equation}
where
\begin{eqnarray}
\Pi_1^{(1)}(k)=\sqrt{k^2}\left(1+\frac{4m^2}{k^2}\right)\log
\left(\frac{1-\sqrt{\frac{4m^2}{k^2}}}{1+\sqrt{\frac{4m^2}{k^2}}}\right)
+4m \,\,,\\
\nonumber \\
\Pi_1^{(2)}(k)=-\frac{4}{\sqrt{k^2}}\log \left(\frac{1-\sqrt{\frac{4m^2}
{k^2}}}{1+\sqrt{\frac{4m^2}{k^2}}}\right)\,\,.
\end{eqnarray}
If we decompose the electromagnetic fields which appear in the integrand of
(47) into the respective real and imaginary parts we see that $\varphi[A]$ is
indeed real. The S-operator in Fock space ${\bf S}[A]$ is then completely
determined.

By means of (12) and (20) we obtain the vacuum-vacuum amplitude
\begin{equation}
(\Omega,{\bf S}\Omega)=Ce^{i\varphi}(\Omega,e^{S_{+-}S_{--}^{-1}b^{\dagger}
d^{\dagger}}\Omega)
=Ce^{i\varphi}\,\,.
\end{equation}
The absolute square
\begin{equation}
|(\Omega,{\bf S}\Omega)|^2=C^2=1-P
\end{equation}
must be equal to one minus the total probability $P$ of pair creation,
\begin{equation}
P={(2\pi)}^2\int\,d^3k\,\hat{P}^{\mu \nu}(k)A_{\mu}(k)A_{\nu}^{*}(k)\,\,,
\end{equation}
since the external field can change the vaccum state only into pair states.
In order to combine the normalization constant $C$ with $e^{i\varphi}$ we
write the former in the exponential form
$$C=\exp\left[2\pi^2\int\,d^3k\,\dots\,\,+O[A^4]\right] \,\,.$$
Hence, from (31)-(35) we get
\begin{flushleft}
$C=\displaystyle{\exp\{-\pi^2\beta\int\,d^3k\,\left[\left(\frac{k^{\mu}k^
{\nu}}{k^2}-g^{\mu \nu}\right)\Pi_2^{(1)}(k^2)+im\varepsilon^{\mu \nu \alpha}
k_{\alpha}\Pi_2^{(2)}(k^2)\right]}$
\end{flushleft}
\begin{equation}
\times A_{\mu}(k)A_{\nu}^{*}(k)\} \,\,,
\end{equation}
where
\begin{eqnarray}
\Pi_2^{(1)}(k^2)=2\sqrt{k^2}\left(1+\frac{4m^2}{k^2}\right)\Theta(k^2-4m^2)
\,\,,\\
\nonumber \\
\Pi_2^{(2)}(k^2)=\frac{8\Theta(k^2-4m^2)}{\sqrt{k^2}}\,\,.
\end{eqnarray}

Finally, taking into account (47)-(49),(50) and (53)-(55), we obtain the
vacuum-vacuum amplitude
\begin{flushleft}
$(\Omega,{\bf S}\Omega)=\displaystyle{\exp\{-i\pi^2\beta\int\,d^3k\,\left
[\left(\frac{k^{\mu}k^{\nu}}{k^2}-g^{\mu \nu}\right)\Pi^{(1)}(k^2)+im
\varepsilon^{\mu \nu \alpha}k_{\alpha}\Pi^{(2)}(k^2)\right]}$
\end{flushleft}
\begin{equation}
\times A_{\mu}(k)A_{\nu}^{*}(k)\} \,\,,
\end{equation}
where
\begin{eqnarray}
\Pi^{(1)}(k^2)=\Pi_1^{(1)}(k^2)-i\Pi_2^{(1)}(k^2) \,\,,
\nonumber \\
\Pi^{(2)}(k^2)=\Pi_1^{(2)}(k^2)-i\Pi_2^{(2)}(k^2) \,\,.
\end{eqnarray}

\newpage

{\large\bf 5.\,Concluding Remarks}
\vspace{1.0cm}

We have considered $QED_3$ in the presence of an external electromagnetic
field $A$ and shown that a unitary scattering operator ${\bf S}$ which
satisfies the local causality condition can be constructed in Fock space.
We have also derived the vacuum-vacuum amplitude and stablished the
connection with vacuum polarization in lowest order of perturbation theory.
In contrast with the four-dimensional case, the vacuum-vacuum amplitude is
ultraviolet finite and exhibits an additional contribution from the
antisymmetric part of the vacuum polarization tensor in (2+1)-dimensional
space-time${}^{[3]}$, which emerges from the topological structure of the
theory.

\vspace{1.0cm}

{\large\bf 6.\,References}
\vspace{1.0cm}

\begin{sf}

\begin{description}
\item[{[1]}] H. P. Seipp, Helv. Phys. Acta {\bf 55} (1982) 1;
\item[{[2]}] G. Scharf, Finite Quantum Electrodynamics, Springer Verlag,
Berlin (1989);
\item[{[3]}] G. Scharf, W. F. Wreszinski, B. M. Pimentel and J. L. Tomazelli,
Ann. of Phys. {\bf 231} (1994), 185;
\item[{[4]}] N. N. Bogoliubov, D. V. Shirkov, Introduction to the Theory of
Quantized Fields, John Wiley, New York (1980).
\end{description}
\end{sf}
\end{document}